\documentclass{aa}
\usepackage{psfig}
\def\kms{{$\rm \, km\, s^{-1}$}}

\begin{document}

\newcommand{\Kkms}{{K~km~s$^{-1}$}\ }
\newcommand{\Htwo}{{H$_{2}$}\ }
\newcommand{\coa}{{$^{12}$CO(1-0)}\ }
\newcommand{\cob}{{$^{12}$CO(2-1)}\ }
\newcommand{\Msun}{{M$_{\odot}$}\ }
\newcommand{\MHtwo}{{M$_{\rm H_2}$}\ }
\newcommand{\x}{{X$_{\rm CO}$}\ }
\newcommand{\xgal}{{X$_{\rm Gal}$}\ }
\newcommand{\xgalv}{{X$_{\rm Gal} = 2.3 \cdot 10^{20}$~cm$^{-2}$~(K~km~s$^{-1}$)$^{-1}$}\ }
\newcommand{\cunit}{{$10^{20}$~cm$^{-2}$~(K~km~s$^{-1}$)$^{-1}$}\ }
\newcommand{\Tmb}{{T$_{\rm mb}$}\ }
\newcommand{\Tsys}{{T$_{\rm sys}$}\ }
\newcommand{\vrel}{{v$_{\rm rel}$}\ }
\newcommand{\Rc}{{R$_{\rm c}$}\ }
\newcommand{\dv}{{$\Delta \rm v$}\ }
\newcommand{\Mvir}{{M$_{ \rm vir}$}\ }
\def\changed{}
\def\changedm{}
\def\kms{{km~s$^{-1}$}}
\def\fmag{\hbox{$.\!\!^{\rm m}$}}
\def\sq{\hbox{\rlap{$\sqcap$}$\sqcup$}}
\def\degr{\hbox{$^\circ$}}
\def\arcmin{\hbox{$^\prime$}}
\def\arcsec{\hbox{$^{\prime\prime}$}}
\def\utw{\smash{\rlap{\lower5pt\hbox{$\sim$}}}}
\def\udtw{\smash{\rlap{\lower6pt\hbox{$\approx$}}}}
\def\fd{\hbox{$.\!\!^{\rm d}$}}
\def\fh{\hbox{$.\!\!^{\rm h}$}}
\def\fm{\hbox{$.\!\!^{\rm m}$}}
\def\fs{\hbox{$.\!\!^{\rm s}$}}
\def\fdg{\hbox{$.\!\!^\circ$}}
\def\farcm{\hbox{$.\mkern-4mu^\prime$}}
\def\farcs{\hbox{$.\!\!^{\prime\prime}$}}
\def\fp{\hbox{$.\!\!^{\scriptscriptstyle\rm p}$}}
\def\err{{$\pm$}}
\hyphenation{paper}
\thesaurus{03(09.04.1; 11.09.1 M82; 11.09.4; 11.19.3; 13.18.1; 13.18.3)}
\title{Cold dust in the starburst galaxy M\,82}
%
%\subtitle{}
%
%
\date{Received 10 March 2000 / Accepted 3 April 2000}  % must be given at the last moment
\author{G. Thuma\inst{1}
 \and N. Neininger\inst{2}
 \and U. Klein\inst{2}
 \and R. Wielebinski\inst{1} }
 \offprints{G. Thuma (gthuma@mpifr-bonn.mpg.de)}
\institute{Max--Planck--Institut f\"ur Radioastronomie,
           Auf dem H\"ugel 69, D--53121 Bonn, Germany
     \and  Radioastronomisches Institut der Universit\"at Bonn,
           Auf dem H\"ugel 71, D--53121 Bonn, Germany
          }
%
% Please don't overrule this command, it will format your titlepage.
\maketitle
\markboth{Cold dust in the starburst galaxy M\,82}{Thuma et\,al.}
\begin{abstract}

The starburst galaxy M\,82 has been mapped at $\lambda$1.2~mm using the
IRAM 30--m telescope and the MPIfR 19--channel bolometer. The influence
of the sidelobe pattern and the $^{12}$CO(2$-$1) line has been carefully
analyzed. Based on this analysis, we conclude that the cold dust does not
extend far into the halo as has been previously claimed. 
\changed{The total mass
of dust in the inner 3~kpc of the galaxy is $\changedm{7.5 \cdot 10^6}$ 
M$_\odot$.
Assuming a solar gas-to-dust ratio, the inferred total mass of gas is 
$\changedm{7.5 \cdot 10^8}$ \Msun.}

Unlike in any other galaxy we find the CO(2$-$1) line emission to be clearly
more extended than the $\lambda$1.2\,mm dust emission.
%This exceptional behaviour is seen for the first time in galaxies.
A likely explanation for this exceptional behaviour,
seen for the first time in galaxies, is heating of molecular
gas by low--energy cosmic rays and soft X--rays in the halo of M\,82.

\end{abstract}

\section{Introduction} \label{sec:intro}

The prototypical starburst galaxy M\,82 is the prime example of a galaxy
in which the violent star formation activity gives rise to the formation
of a bipolar outflow and an associated extended halo, which is visible in
various regimes of the electro--magnetic spectrum. Outflowing material was
first evident in the H$\alpha$ light (Lynds \changed{ \& } Sandage 1963, McCarthy et
al. 1987, Bland \changed{ \& } Tully 1988). The kinematics of the outflowing material
and the geometry of the cone have been thoroughly worked out by McKeith et
al. (1995). The recent detection of H$\alpha$ emission $\sim$11~kpc away from
the plane of M\,82 (Devine \changed{ \& } Bally 1999) determines the cap of this
outflow. The whole scenario is further corroborated by the existence of
vertical magnetic field lines (Reuter et\,al. 1994), along which relativistic
particles partake in the outflow, thus forming a radio halo (Seaquist \&
Odegard 1991). This halo exhibits a filamentary structure away from the
plane (Reuter et\,al. 1992). Hot gas has been also found in some kind of halo
which extends for several arcminutes along the minor axis
(Fabbiano 1988, Schaaf et\,al. 1989, Bregman et\,al. 1995).
This X--ray emission correlates well with the H$\alpha$
if observed with sufficiently high angular resolution (Watson et\,al. 1984).

In this context, the search for neutral gas and dust away from the plane
of M\,82 has seen a number of attempts. Owing to the close interaction
between M\,81 and M\,82, neutral hydrogen is seen enveloping M\,82, but
whether this gas stems from the M\,81 or has been expelled from M\,82
is still a matter of debate (Yun et\,al. 1994). Molecular gas associated
with the dusty filaments outside the plane of M\,82 has been reported by
Stark \changed{ \& } Carlson (1984) and has been observed by Sofue et\,al. (1992)
out to a projected distance of $\pm$2~kpc. Clear evidence for the existence
of scattering dust away from the plane of M\,82 comes from measurements of
optical polarization (Bingham et\,al. 1976, Notni et\,al. 1981,
Neininger et\,al. 1990).

Owing to its high luminosity in all spectral bands, M\,82 has been a
prime target for first--light experiments in the mm-- and submm--regime
(Elias et\,al. 1978, Jura et\,al. 1978). With the improvement of
bolometric measurements detailed studies of the distribution of the cold
dust in this galaxy have become feasible; thus the recent past has seen
an increasing number of such investigations ($\lambda$2~mm: Kuno \changed{ \& } Matsuo 1997;
$\lambda$1~mm: Thronson et\,al. 1989, Hughes et\,al. 1990, Kr\"ugel et\,al. 1990a,
Alton et\,al. 1999; $\lambda$0.5~mm: Jaffe et\,al. 1984, Smith et\,al. 1990, Alton
et\,al. 1999). Carlstrom \changed{ \& } Kronberg (1991) have shown that at $\lambda$3~mm
thermal free--free emission still dominates the overall spectrum.

Apart from these dedicated bolometric measurements, continuum maps at
millimeter wavelengths have been produced as by-products in various
observations of spectral lines (e.g. Neininger et\,al. 1998).
As a result of the discoveries of a galactic wind and halo in M\,82, the more
recent  bolometric measurements aimed at detecting a dust halo at mm wavelengths.
CO emission away from the plane was reported by Stark \changed{ \& } Carlson (1984), Nakai
et\,al. (1986) and by Sofue et\,al. (1992). An outflow of molecular gas was claimed by Nakai et\,al.
(1987). A dust continuum halo was first mentioned by Hughes et\,al. (1990).
Observations with higher resolution (Kuno \changed{ \& } Matsuo 1997) suggest emission at
$\lambda$2~mm out to 400~pc from the plane. Recently Alton et\,al. (1999) reported
a dust outflow from the central region of M\,82, based on their sub-mm images.

Here we report observations of M\,82 in the $\lambda 1.2$ mm continuum
performed with the MPIfR 19-channel bolometer at the IRAM 30-m telescope.
In Sect.~\ref{sec:obse} we describe the observations and data analysis,
with consideration of the possible influence of the error beam. In
Sect.~\ref{sec:result} the distribution of the cold dust in M\,82 will
be presented, with a discussion conducted in Sect.~\ref{sec:discuss},
along with a comparison with other published measurements. In
Sect.~\ref{sec:sum} we give a short summary of our results.

\section{Observations and data reduction} \label{sec:obse}

\subsection{Observations} \label{sec:obse:obse}

We used the 19-channel bolometer of the MPIfR (Kreysa et\,al. 1993) in the
Nasmyth focus of the IRAM 30-m telescope during the spring session in 1997.
The weather conditions were fair during our night-time runs, with rather
stable zenith opacities between 0.2 and 0.3 at 230~GHz.

The 19 channels of the bolometer are arranged in a closely packed hexagonal
array, with beamsizes of 11\arcsec\ FWHM and spacings of 20\arcsec. The
calibration was performed by mapping Uranus every morning. The pointing and
focus checks were made at regular intervals during the observations using
nearby  quasars (in particular 1308+326). The map was centered on the
near--infrared  peak at $\alpha_{50} = 09^{\rm h} 51^{\rm m} 43\farcs4$,
$\Delta_{50} =  69\degr 55\arcmin 01\arcsec$ (Joy et\,al. 1987). The data were
taken in the  standard mapping mode where the field is scanned at a constant
speed of  4\arcsec\ per second in azimuth, with subsequent scans being spaced
by  4\arcsec\ in elevation. The map size was chosen large enough (320\arcsec
$\times$ 270\arcsec) to assure a good baseline determination in the presence
of extended emission. We thus made sure that the field was mapped sufficiently
far out perpendicular to the major axis of M\,82, such as to trace the full
extent of any dust component in the halo. While scanning the subreflector was
wobbled at a 2\,Hz rate with an amplitude of 45\arcsec, which is small enough
to ensure a stable beam pattern. Due to the small wobbler throw the OFF position
was not necessarily free of emission, but using the algorithm of Emerson et
al. (1979) this should have no effect on the resulting map. We also mapped
Mars once and the point source 3C\,273 twice each in order to obtain a
detailed beam pattern. In addition, more maps of Mars and other calibration
sources were used to properly check the calibration.

\subsection{Standard data reduction} \label{sec:obse:standard}

The whole field was covered six times. For each coverage we performed the
data reduction separately using NIC (GILDAS software package), starting with
the subtraction of a third--order baseline.

The scanning mode described in the previous section leads to maps which contain
a superposition of a positive and a negative image of the source (double-beam
maps). In principle the deconvolution can be done by dividing the Fourier
transform (FT) of the measured intensity distribution by the FT of the wobble
function and transforming back to image space. As the FT of the wobble
function is a sine wave with zeroes at the origin and at harmonics of the
inverse wobbler throw, \changed{it would cause} problems at these spatial
frequencies. To avoid these problems we used the algorithm of Emerson et\,al.
(1979) to restore the double--beam maps into equivalent single--beam maps.

Thereafter we combined the 19 channels and calibrated the resulting map
with respect to Mars and Uranus (assuming brightness temperatures
of 198.5~K and 97.2~K respectively). In a last step the four best coverages
were averaged and regridded onto equatorial (B1950) coordinates. (Two coverages were ignored
because of bad S/N ratios.) This final map has a sensitivity of 5~mJy per beam
and a angular resolution of 12\arcsec\ (HPBW).

\subsection{Influence of the error beam} \label{sec:obse:errorbeam}

In order to check the influence of the error beam on the extended emission we
applied a CLEAN algorithm developed for single--dish maps (Klein \changed{ \& } Mack 1995)
to each coverage using a radially symmetric antenna pattern. This pattern was
obtained in the following way: A field of 4\farcm5 $\times$ 3\farcm5 size was
mapped centred on 3C\,273. This antenna pattern was then
rotated in small steps ($5\degr$), and these individual (rotated) maps were
averaged to yield a nearly radially symmetric pattern. The subtraction of the
CLEANed map from the original one showed that the error beam contributes less
than 25\% of the rms noise level to the extended emission ($\ge20$\arcsec\ away
from the central position).

Even the sidelobes can be ignored in our case, because at $\lambda$ 1.2\,mm they are of the
order of 2\%, corresponding to roughly 5\,mJy/beam in the M82 maps. Averaging
four coverages with different parallactic angles (and accordingly different
positions of the sidelobes) the effect of the antenna pattern on the extended
emission falls well below the rms noise of our final map.

This extra analysis make sure that the extended emission seen in our map is real
and not affected by the error beam.

\subsection{Correction for CO emission} \label{sec:obse:cocorr}

\begin{figure}[h]
\begin{center}
\psfig{file=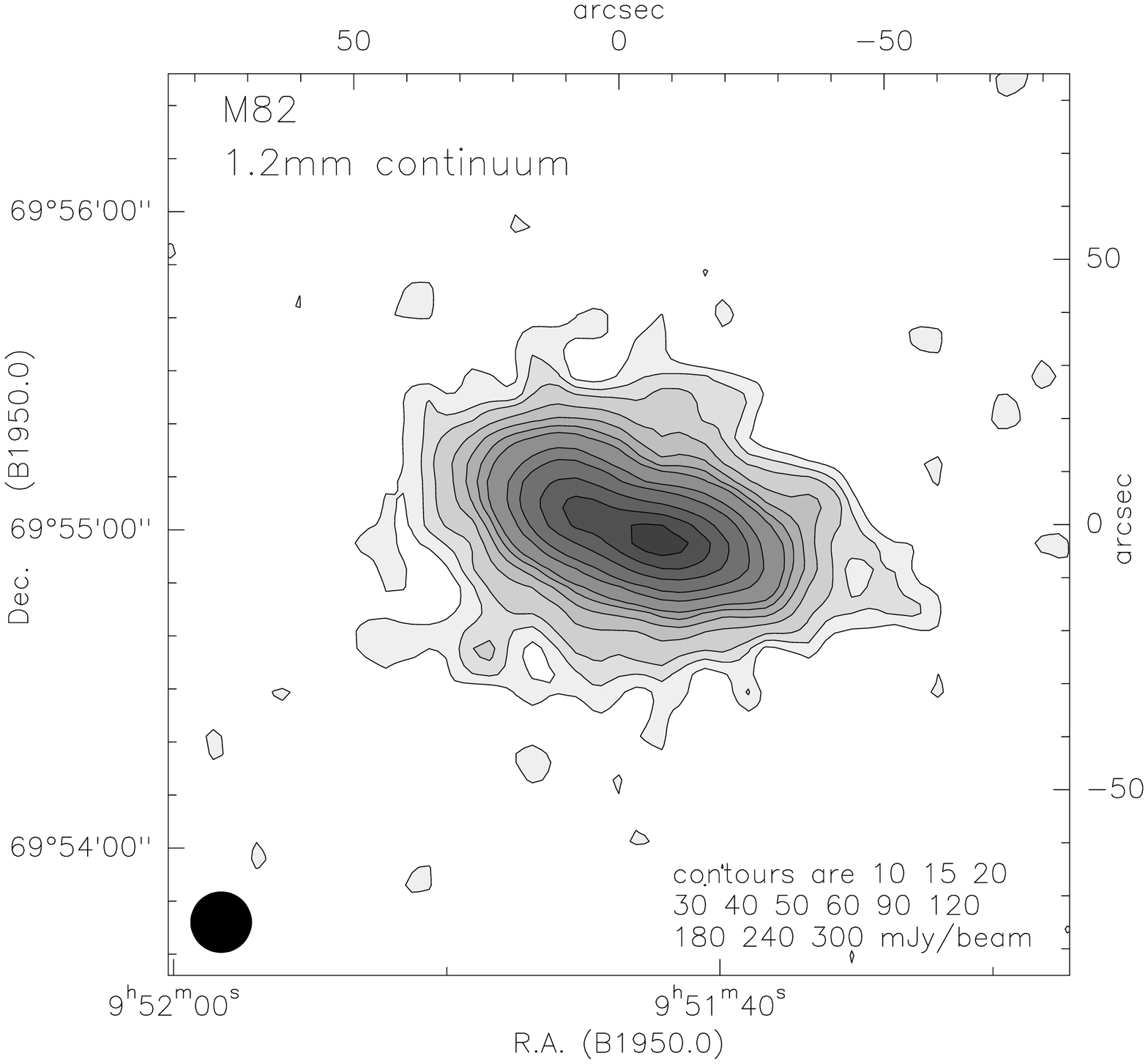,width=8.5cm,angle=270}

\vspace*{1cm}

\psfig{file=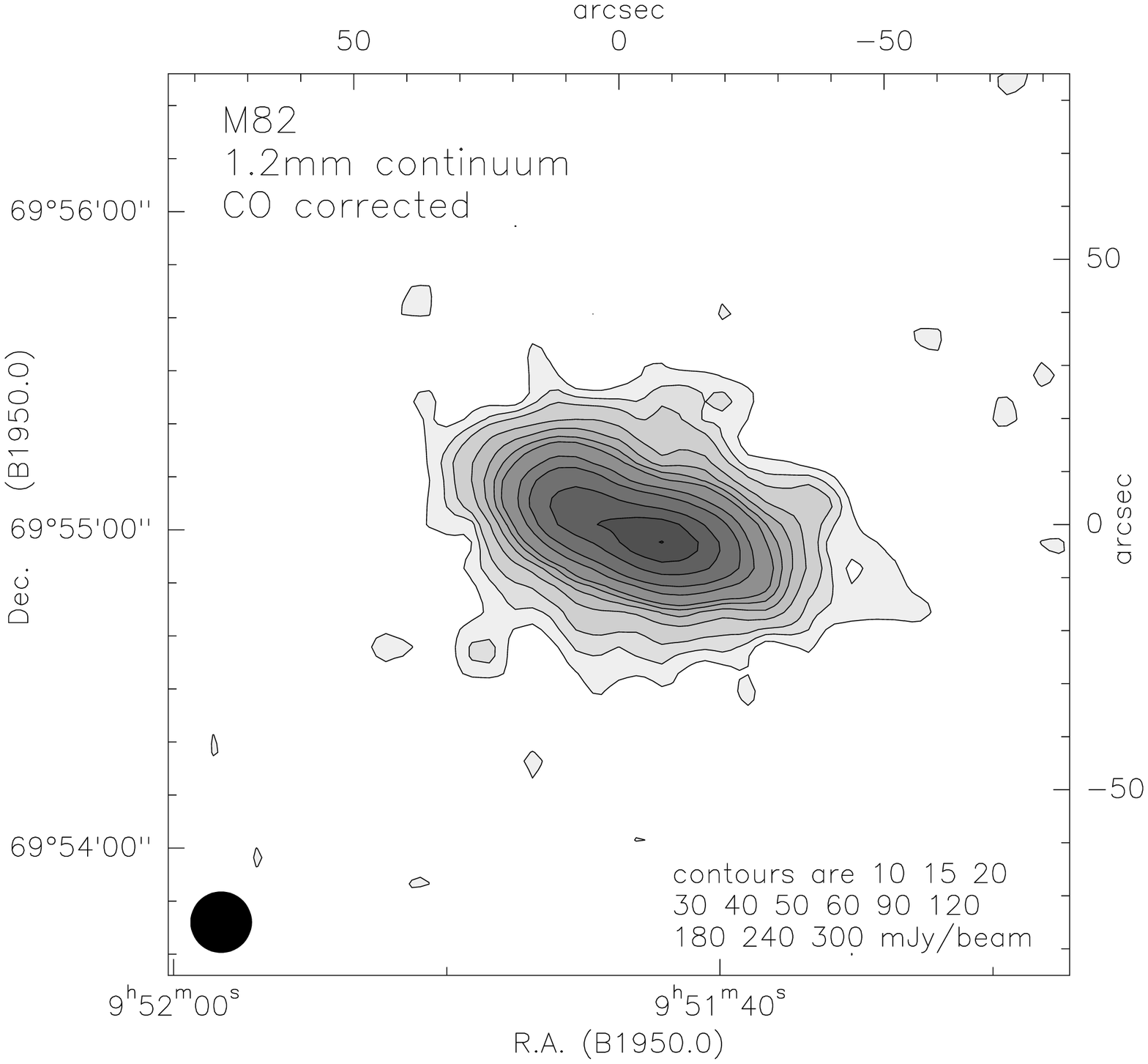,width=8.5cm,angle=270}
\caption{Maps of M\,82 at $\lambda$1.2~mm. The upper map shows the continuum
emission after standard data reduction. The lower one shows the continuum
emission, with the CO line emission subtracted. The contours are the same in
both images starting at the 2$\sigma$ (10~mJy/beam) level. The beamsize of
the IRAM 30-m telescope is shown in the lower left corner of each image.}
\label{fig:mitohne}
\end{center}
\end{figure}

Using the 19-channel bolometer, with its central frequency of 240~GHz and its
effective bandwidth of $\Delta\nu_{bol}\approx$~80~GHz, the
$^{12}$CO(2$-$1)~line at 230~GHz might
noticeably contribute to the total intensity in the bandpass. To get a reliable
information on the distribution of the cold dust, we applied the following
correction to our continuum map:

We used a $3' \times 3'$ map of M\,82 made with the same telescope in the
$^{12}$CO(2-1) line (obtained to provide the zero spacings for an
investigation with the Plateau de Bure interferometer,
Wei\ss{} et\,al.in prep.).
This map was measured in the on--the--fly mode. The on--the--fly technique
ensures a uniform sampling of the emission over the whole field with good
signal-to-noise ratio. Since we knew from our continuum map that the effects
of the error beam can be neglected at a frequency of 240~GHz, we did not
apply any error beam correction to the CO map.

Integrating over the effective bandwidth of the bolometer, the total continuum
flux in the beam at the peak position is given by $$\rm
S^{bol}=\Delta\nu_{bol}\cdot \rm S^{bol}_{240}=2.6\cdot 10^{-16}
\,\rm Wm^{-2}$$ where $\rm{S}^{\rm{bol}}_{240} = 330$~mJy is the peak flux at
240~GHz.

The integrated intensity of the $^{12}$CO(2$-$1) line at the same position
$\int{\rm{T}_{\rm{mb}}\,\rm{dv}}=625\,\rm{K\,km\,s}^{-1}$
was taken from our CO map. Expressing this in terms of frequency instead
of velocity, the integrated intensity reads
\mbox{$\int{\rm{T}_{\rm{mb}}\,\rm{d\nu}}=500\,\rm{K\,MHz}$.}

To obtain the total flux in the $^{12}$CO(2$-$1) line we used the
Rayleigh--Jeans Approximation and assumed that
T$_{\rm{mb}}\approx \rm{T}_{\rm{ex}}$:
$$\rm S^{CO}=\Delta\nu_{\rm{CO}}\cdot\Omega\cdot
\frac{2\,\rm{kT_{\rm{ex}}}}{c^2}\cdot\nu^2\approx
\Omega\cdot\frac{2\,\rm{k}\nu^2}{c^2}\cdot
\int{\rm{T}_{\rm{mb}}\,\rm{d\nu}}$$
With the beam solid angle $\Omega = 4.5\cdot10^{-9}$sr and the integrated
$^{12}$CO(2$-$1)
intensity from above we get \mbox{$\rm S^{CO}=3.6\cdot 10^{-17}\,\rm Wm^{-2}$.}
Thus, $\rm{S}^{\rm{CO}}/\rm{S}^{\rm{bol}}\approx 0.14$ at the position of the
continuum peak.

If we assume that this value holds for the whole galaxy,
which of course is a simplification, we can obtain a CO--corrected
continuum map by subtracting the appropriately scaled CO map from our
continuum map. The uncorrected and the corrected map are shown in Fig.
\ref{fig:mitohne}.

\section{Results} \label{sec:result}

\subsection{Overall distribution of the dust emission} \label{sec:result:shape}

\begin{figure*}
\begin{center}
\vbox{
\psfig{file=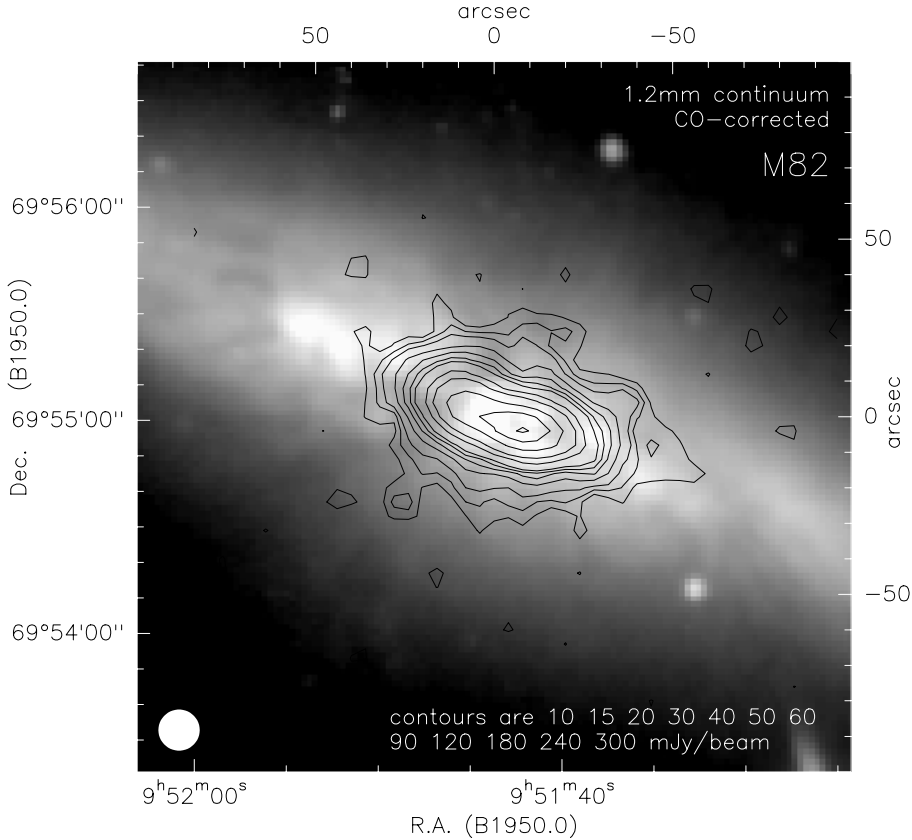,width=12cm,angle=270}\vspace{-3.5cm}}
\hfill\parbox[b]{5.5cm}{
\caption{CO--corrected map of M\,82 at $\lambda$1.2~mm overlayed on a
R-band image made with the 1.2-m telescope on Calar Alto.
The contours start at the 2$\sigma$ (10~mJy/beam) level.
The beamsize of the IRAM 30-m telescope is given in the lower left corner.}
\label{fig:cont}}
\end{center}
\end{figure*}

Fig.~\ref{fig:cont} shows the CO-corrected continuum map of M\,82 at
$\lambda$1.2\,mm, superimposed onto an optical image. At $\lambda$1.2\,mm
M\,82 has a more or less oval shape (ellipticity e~$\approx$~0.6),
with the major axis parallel to the galactic plane (molecular gas).
This corresponds to a position angle of approximately 70\degr.
The maximum of the continuum emission is located at
$\alpha_{50} = 9^{\rm h}51^{\rm m}41^ {\rm s}$, $\Delta_{50} =
69$\degr54\arcmin55\arcsec. Within the errors this position is identical to
the peaks of the distributions at $\lambda$800\,$\mu$m and $\lambda$1.1\,mm
(Hughes et\,al. 1990). Besides this regular shape there are some spur--like
features extending
above and below the galactic plane, which might be associated with the
outflow seen at optical and IR wavelengths.

\subsection{Major-- and minor--axis profiles} \label{sec:result:profile}

\begin{figure}
\begin{center}
\psfig{file=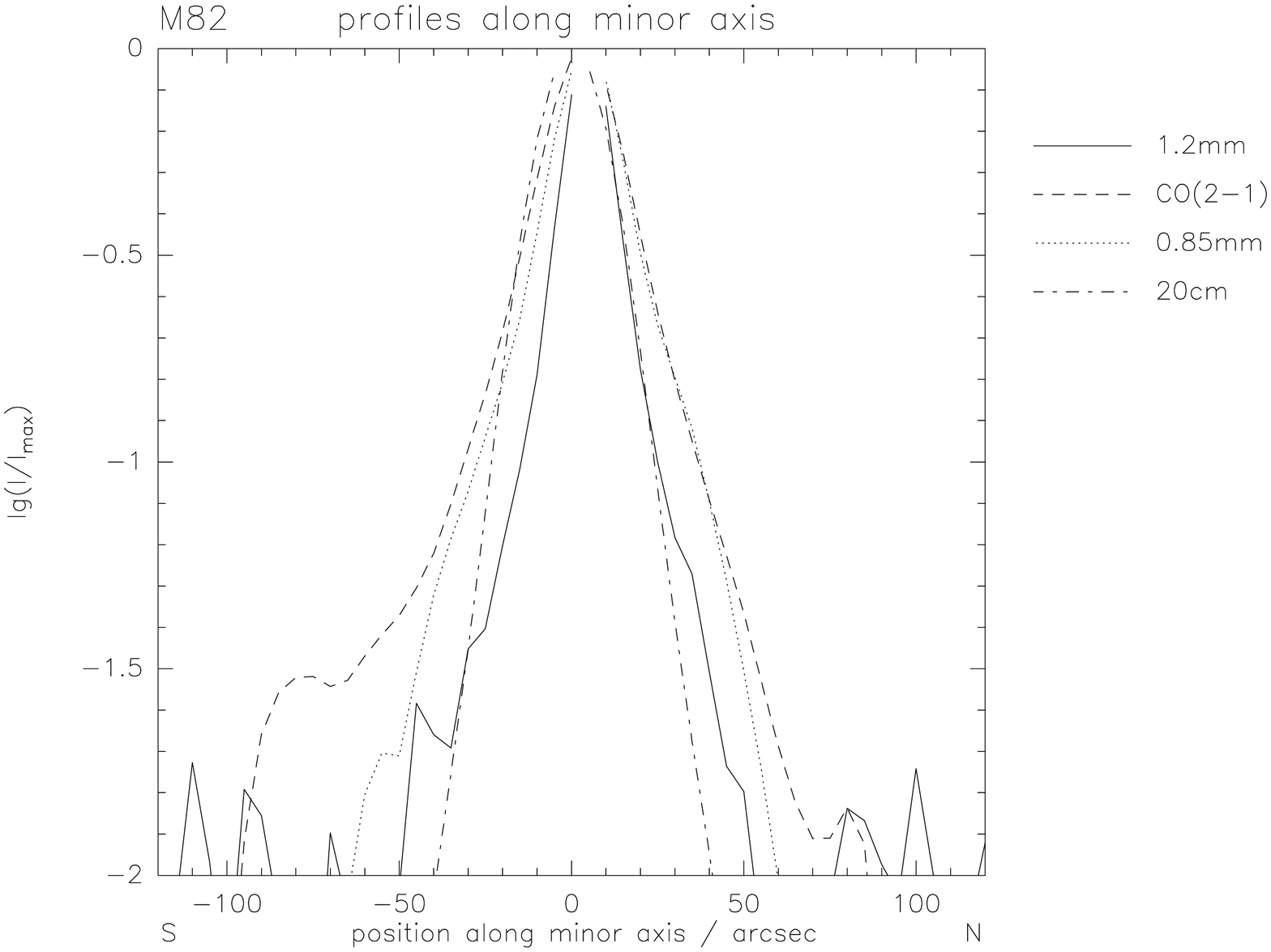,width=8.5cm,angle=270}

\vspace{1cm}

\psfig{file=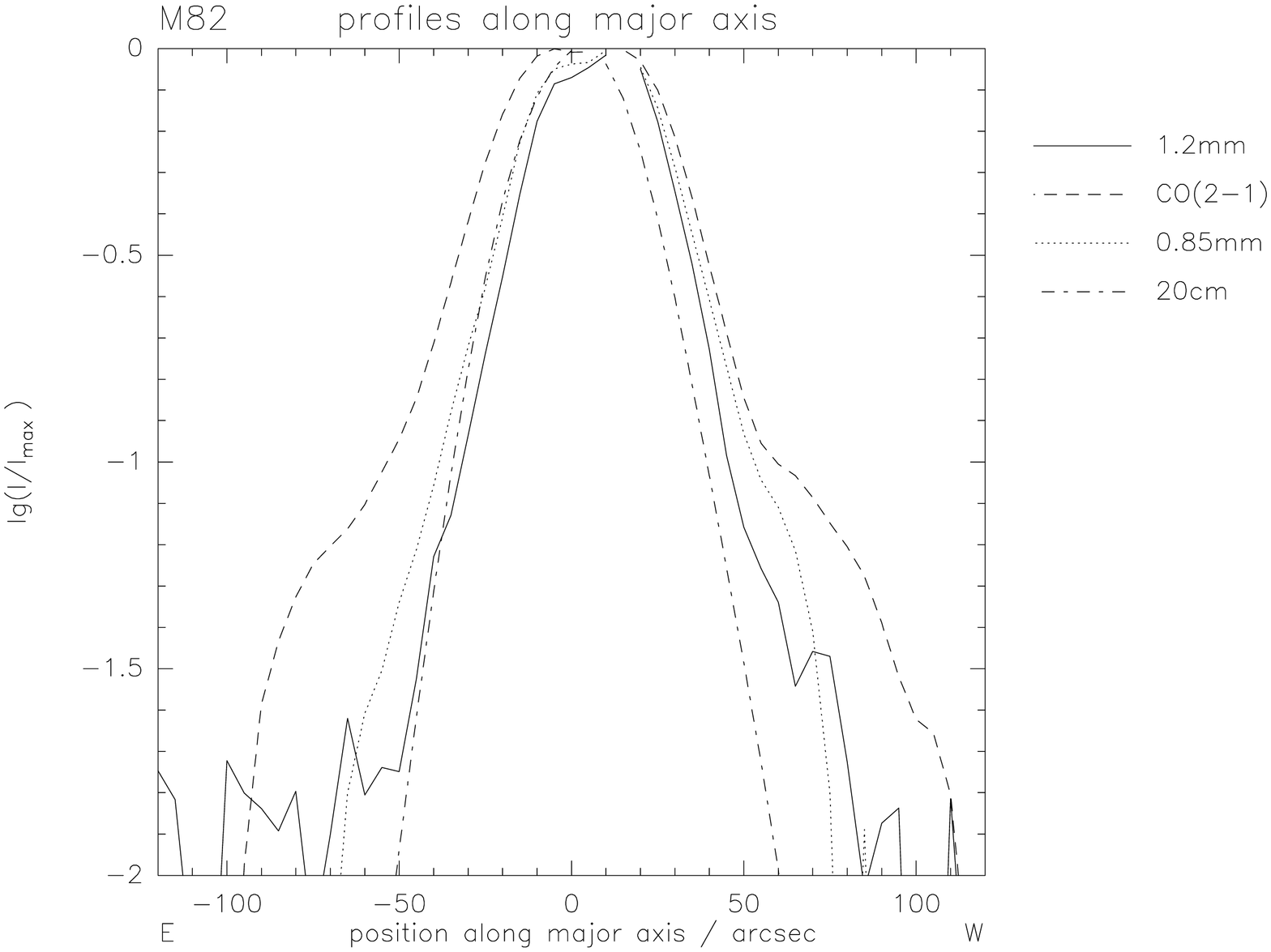,width=8.5cm,angle=270}
\caption{Profiles along the minor and major axis of M\,82. For each direction
the $\lambda$1.2~mm continuum emission (this paper), the integrated CO(2$-$1)
line intensity (this paper), the $\lambda$850~$\mu$m continuum emission
(Alton et\,al. 1999) and the radio continuum emission at $\lambda$20~cm (Reuter
et\,al. 1992) are shown. All data are smoothed to a beam of 15\arcsec ~HPBW. }

\label{fig:slice}
\end{center}
\end{figure}

The contours in Fig.~\ref{fig:cont} indicate some extended emission also along
the minor axis of the  galaxy, but at first glance the extent in this
direction is not as large as suggested by the maps of Kuno \changed{ \& } Matsuo (1997)
and Alton et\,al. (1999). We have therefore computed the intensity as a
function of distance from the galactic center along the minor (major) axis,
averaging over some 30\arcsec\ in the major-axis (minor-axis) direction.
These profiles are displayed in Fig.~\ref{fig:slice}. For comparison the
$^{12}$CO(2$-$1) line intensity, the 850~$\mu$m (Alton et\,al. 1999) and 20~cm
continuum profiles are shown, too.

The $\lambda$1.2~mm emission shows the most concentrated distribution, significantly
smaller than that at 850~$\mu$m. Especially the scale height along the minor
axis is very small with respect to the other wavelengths and the molecular
emission.

Contrary to the expectation the molecular emission is the most
extended component. Possible explanations for this phenomenon are discussed
in Sect. \ref{sec:discuss:co}.

\subsection{Integrated flux density} \label{sec:result:spec}

\begin{figure*}
\begin{center}
\vbox{
%\hspace*{0.5cm}
\psfig{file=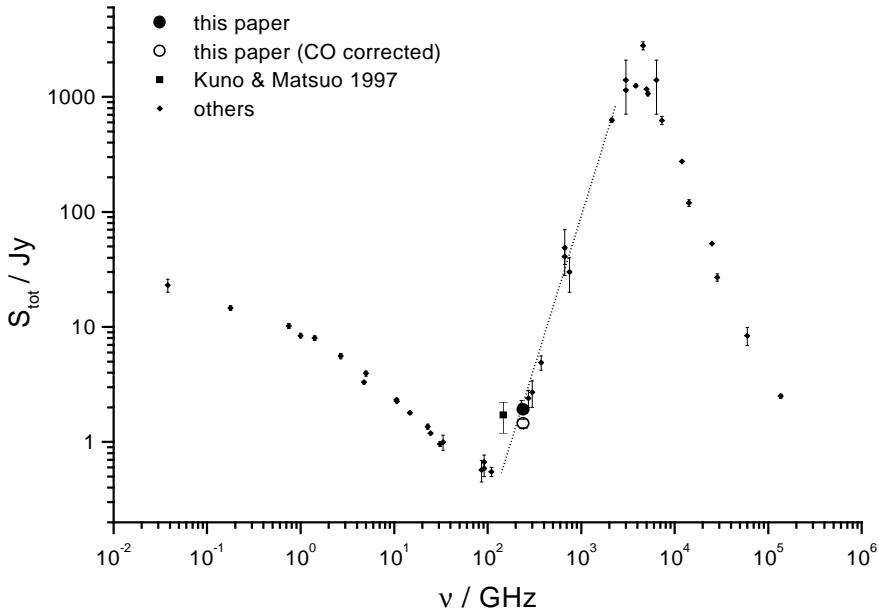,width=12cm}\vspace{-4.3cm}}
\hfill\parbox[b]{5.5cm}{
\caption{Radio-to-IR spectrum of M\,82. Our new value is indicated
by the filled circle, the CO--corrected value by the open circle, and the
115~GHz point of Kuno \& Matsuo 1997 by the filled square. The slope of the
spectrum in the submm wavelength range is indicated by the dotted line. The
error bars of our data points are smaller than the used symbols.}
\label{fig:spektrum}}
\end{center}
\end{figure*}

\begin{table}
\begin{center}
\caption{Flux densities of M\,82 from radio through IR wavelengths.}
\label{tab:spectrum}
\begin{tabular}{|c|c|l|}
\hline
$\nu$/GHz & S$_{\nu}$/Jy & Ref. \\
\hline
0.038	&23\err 3		&Kellermann et\,al. 1969\\
0.178	&15.3\err 0.7	&Kellermann et\,al. 1969\\
0.75	&10.7\err 0.5	&Kellermann et\,al. 1969\\
1		&8.6\err 0.4	&Kellermann et\,al. 1969\\
1.415	&8.2\err 0.4	&Hummel 1980\\
2.695	&5.7\err 0.3	&Kellermann et\,al. 1969\\
4.75	&3.3\err 0.1	&Klein et\,al. 1988\\
5		&3.9\err 0.2	&Kellermann et\,al. 1969\\
10.63	&2.3\err 0.08	&Doherty et\,al. 1969\\
10.7	&2.25\err 0.06	&Klein et\,al. 1988\\
14.7	&1.79\err0.04	&Klein et\,al. 1988\\
22.8	&1.36\err 0.07	&Kronberg et\,al. 1979\\
24.5	&1.19\err 0.03	&Klein et\,al. 1988\\
31		&0.96\err 0.05	&Kellermann et\,al. 1971\\
33.3	&1.0\err 0.2	&Klein et\,al. 1988\\
86		&0.6\err 0.1	&Kellermann et\,al. 1971\\
92		&0.7\err 0.1	&Seaquist et\,al. 1996\\
92		&0.59\err 0.09	&Carlstrom \& Kronberg 1991\\
110		&0.55\err 0.05	&Neininger et\,al. 1998\\
115		&1.7\err 0.5	&Kuno \& Matsuo 1997\\
230		&2.1\err 0.2	&Kr\"ugel et\,al. 1990a\\
\textbf{230}		&\textbf{1.91\err 0.13}	&\textbf{this paper}\\
273		&2.4\err 0.4	&Hughes et\,al. 1990\\
300		&2.7\err 0.7	&Elias et\,al. 1978\\
375		&4.9\err 0.7	&Hughes et\,al. 1990\\
667		&41\err 6		&Hughes et\,al. 1994\\
667 	&49\err 21		&Smith et\,al. 1990\\
750		&30\err 10		&Jaffe et\,al. 1984\\
2130	&630\err 24		&Telesco \& Harper 1980\\
3000	&1400\err 690	&Harper \& Low 1973\\
3000	&1350\err200	&Rice et\,al. 1988\\
3850	&1255\err 34	&Telesco \& Harper 1980\\
4615	&2800\err 220	&Harper \& Low 1973\\
5000	&1270\err190	&Rice et\,al. 1988\\
5170	&1066\err 44	&Telesco \& Harper 1980\\
6383	&1400\err 690	&Harper \& Low 1973\\
7320	&635\err 51		&Telesco \& Harper 1980\\
12000	&285\err43			&Rice et\,al. 1988\\
14286	&120\err 8		&Rieke \& Low 1972\\
25000	&67\err10		&Rice et\,al. 1988\\
28571	&27\err 27		&Rieke \& Low 1972\\
60000	&8.4\err 1.5	&Rieke \& Low 1972\\
136364	&2.5\err 0.1	&Rieke \& Low 1972\\
\hline
\end{tabular}
\end{center}
\end{table}

Integrating over the continuum map we obtain a total flux density of S$_{240}
= 1.91 \pm 0.13$~Jy. This is lower than the value measured by Kr\"ugel et
al. (1990a), but their flux density was much more uncertain, owing to the
lower sensitivity of their 1-channel bolometer.

\changed{Fig. \ref{fig:spektrum} shows a radio-to-IR spectrum of M\,82 with the 
individual values listed in Tab. \ref{tab:spectrum}.} Our new 240~GHz
measurement is completely consistent with the slope of the spectrum at this
frequency (indicated by the dotted line), whereas  the value derived by
Kuno~\&~Matsuo\- (1997) at 157~GHz lies much too high with respect to all
adjacent data points.

The CO--corrected flux density is significantly lower than the flux density
in our original map. This demonstrates that a line correction for continuum
measurements
in the relevant bands is indispensible.

The total flux can be turned into a
gas mass via 
$$
\rm M_{\rm g} 
= 
\frac{\rm S_{240} \cdot \rm D^2}{\rm B(\rm T_{\rm d})\cdot\kappa'_{240}}
\cdot\frac{\rm M_{\rm g}}{\rm M_{\rm d}}. 
$$
With a distance of D = 3.25~Mpc, and
using the same para\-meters as Kr\"ugel et\,al. (1990a), i.e. a dust
temperature of T$_{\rm d}$ = 30~K (Chini et\,al. 1989) 
and a \changed{dust absorption coefficient
$\changedm{\kappa_{240}=\kappa'_{240}\cdot\frac{\rm M_{\rm d}}{\rm M_{\rm g}}
=0.003 \;\rm cm^2\rm g^{-1}}$ (Kr\"ugel et\,al. 1990b)},
we derive a gas mass of \hbox{$\rm M_g = 7.5 \cdot 10^8\,M_{\odot}$.} 
If we adopted the
Draine \& Lee value $\kappa_{240}=0.005 \;\rm cm^2\rm g^{-1}$ (Draine \& Lee
1984) the resulting gas mass would be somewhat lower ($\rm M_g = 4.5 \cdot
10^8\,M_{\odot}$).

In this calculation we assumed a gas-to-dust ratio of
approximately M$_{\rm g}$/M$_{\rm d} \approx \changed{100}$.
The corresponding dust mass is \changed{M$\changedm{_{\rm d} = 7.5\cdot 
10^6\,M_{\odot}}$.}

\section{Discussion} \label{sec:discuss}

\subsection{A dust halo?} \label{sec:discuss:conti}

The $\lambda$2\,mm map (Kuno \& Matsuo 1997) shows continuum emission with a
box--like shape, indicating a dust halo around M\,82. Although the emission at
$\lambda$1.2\,mm and $\lambda$2\,mm should be similar, we do not see any
/*-evidence for a significant halo contribution to the cold dust.
This difference could be due to the error beam of the Nobeyama 45-m telescope.

Effects of the error beam and the antenna pattern have been investigated and
largely removed from
our map (see Sect. \ref{sec:obse:errorbeam}), while this is not mentioned in
case
of the Nobeyama measurements. The total flux density in the Nobeyama map is
1.7\,Jy, which is much more than expected at this wavelength
(see Fig.~\ref{fig:spektrum}\changed{)}. This extra flux
might be due to emission entering through sidelobes and the error beam.

The 850\,$\mu$m map made by Alton et\,al. (1999) with the JCMT shows continuum
emission  at the 50\,mJy level up to 40\arcsec\ above the galactic plane. This
translates to  15\,mJy at $\lambda$1.2\,mm, corresponding to the second
contour (3$\sigma$) in our  map. Considering only the overall extent of the
emission, the JCMT map is consistent with ours. The different shapes might be
due to problems with the baseline subtraction, because the area mapped by
Alton et\,al. is only slightly larger than the extent of the continuum emission.

\subsection{Comparison of dust and CO} \label{sec:discuss:co}

\begin{figure}
\begin{center}
\psfig{file=h1816f1b.eps,width=8.5cm,angle=270}

\vspace{1cm}

\psfig{file=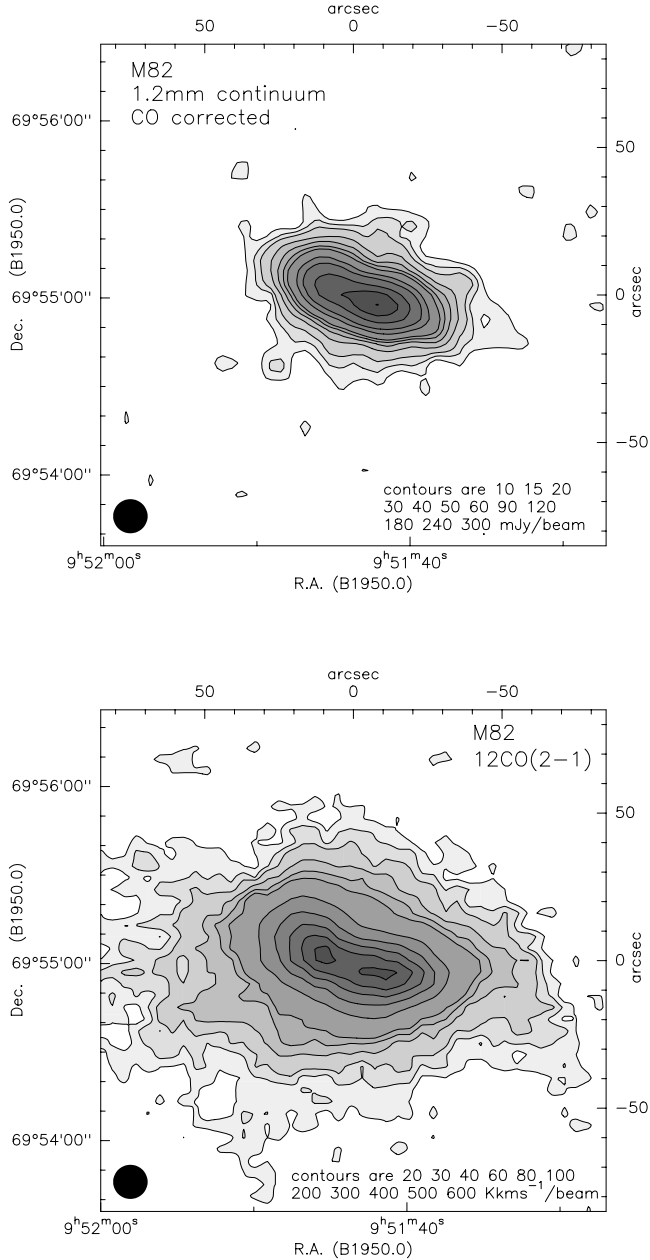,width=8.5cm,angle=270}
\caption{Comparison of the $\lambda$1.2~mm continuum and the CO(2$-$1) line
emission. The upper image shows the CO--corrected $\lambda$1.2~mm
continuum map of M\,82. Contours start at 2$\sigma$ (10~mJy/beam). The lower
image shows the integrated CO(2$-$1) line intensity. Contours start at 3$\sigma$
(20~K\,km\,s$^{-1}$/beam). The beam size of the IRAM 30-m telescope is given in the lower left
corner of each image.}
\label{fig:vergleich}
\end{center}
\end{figure}

In non--active galaxies like NGC\,4565 (Neininger et\,al. 1996) and
NGC\,5907 (Dumke et\,al. 1997) the molecular line emission is more concentrated
towards the central region than the thermal dust emission.
As can be seen from Fig.~\ref{fig:slice} and~\ref{fig:vergleich} the galaxy
M\,82 exhibits
the opposite behaviour. The CO emission appears to be much more extended
than the continuum. Since the maps in Fig. \ref{fig:vergleich} are made with
the same telescope at the same frequency, the difference can not be
due to different beam patterns and error beams.

\changed{The dust is heated by the UV radiation field, which is concentrated
towards the dense central region, because the UV photons can hardly escape 
without being absorbed. Only a small fraction of the UV photons makes its way
out to kpc distances from the galactic plane, where it might contribute to 
the observed polarized radiation. }

The molecular gas can be excited by the radiation
field, but also by low--energy cosmic rays and by soft X--ray photons 
(see e.g. Glassgold \changed{ \& } Langer 1973)\changed{, which have a 
much larger scale height than the UV photons (Shopbell \& Bland-Hawthorn 1998)}.

Fig.~\ref{fig:slice} shows that the synchrotron intensity at $\lambda$20~cm
drops faster with distance from the galaxy plane than the CO and even the 
dust emission.
However, at low radio frequencies there may be a significant radio halo
(low--energy cosmic rays have longer lifetimes), which is already indicated by
the 327--MHz map of Reuter et\,al. (1992). Hence, there may be abundant cosmic
rays for heating. Such cosmic rays (E $\la$100~MeV) remain, however, invisible
in the radio window, since even in the strong ($\rm{B}\approx 10
\ldots 50~\mu$G) magnetic field of M\,82 they produce synchrotron
radiation at
frequencies below about 10~MHz. Soft X--rays are also seen far out of the
plane of M\,82 (e.g. Bregman et\,al. 1995). These circumstances \changed{altogether}
provide the likely heating sources for any molecular material transported out
of the disk of M\,82 into the halo, rendering the CO ''overluminous''. This is
also indicated by the low conversion ratio of CO line intensity to molecular
gas column density, X$_{\rm CO} = \rm N_{\rm H_2}/\rm I_{\rm CO}$ derived by
Smith et\,al. (1991). The existence of a soft X--ray and a low--frequency radio
halo around M\,82 readily explains the larger extent of the CO emission as
compared to the cold dust.

\section{Summary} \label{sec:sum}

In this paper we presented a new high--sensitivity map of the continuum
emission at $\lambda$1.2\,mm in the starburst galaxy M\,82. To get a reliable
map of the extended thermal dust emission we applied a CLEAN algorithm to
our map (error beam correction) and subsequently subtracted the
$^{12}$CO(2-1) line
emission. Besides this the contribution of the sidelobes of the IRAM 30-m
telescope has been estimated.

The total mass of dust in the inner 3~kpc of the galaxy derived from the
integrated $\lambda$1.2\,mm flux density is $5.0 \cdot 10^6$ M$_\odot$,
the inferred total mass of gas is $7.5 \cdot 10^8$ \Msun.

Although the continuum emission is not confined to the galactic disk, our map
rules out the existence of any pronounced halo of cold dust around M\,82. We
find the CO(2$-$1) line emission to be clearly more extended than the
$\lambda$1.2\,mm dust emission. This unexpected observational result now
waits for a conclusive theoretical explanation.

\acknowledgements We thank Dr. A. Sievers from IRAM (Granada) for his
support with additional calibration. We are very grateful to Dr. P.B.
Alton for making his JCMT data available to us.

\end{document}